\documentclass[fleqn,twoside]{article}
\usepackage{espcrc2}

\readRCS
$Id: espcrc2.tex,v 1.2 2004/02/24 11:22:11 spepping Exp $
\usepackage{epsfig}

\newcommand{\AmS}{{\protect\the\textfont2
  A\kern-.1667em\lower.5ex\hbox{M}\kern-.125emS}}

\newcommand{\nn}{\nonumber}
\newcommand{\als}{\alpha_{\rm s}}
\newcommand{\ra}{\rightarrow}
\newcommand{\ep}{\epsilon}

\title{
\vspace*{-18mm} \rightline{ {\normalsize{DESY 06--150}}}
\vspace*{-2mm}  \rightline{\normalsize{September 2006}}
\vspace*{+6mm}
NNLO splitting and coefficient functions with time-like kinematics%
{\thanks{Presented by A.M. at {\it{Loops and Legs in Quantum Field
Theory}}, 23--28 April 2006, Eisenach (Germany).}}%
}

\author{A. Mitov\address[DESYZ]{DESY, Platanenallee 6,
D--15735, Zeuthen, Germany}, S. Moch\addressmark[DESYZ]
and
A. Vogt\address[Durham]{IPPP, Department of Physics, University of Durham,
South Road, Durham DH1 3LE, United Kingdom}
\address[Liverpool]{Department of Mathematical Sciences, University of Liverpool, Liverpool L69 3BX, United Kingdom}
\thanks{Address after 1 September 2006.}}

\begin{document}

\begin{abstract}
We discuss recent results on the three-loop (next-to-next-to-leading order,
NNLO) time-like splitting functions of QCD and the two-loop (NNLO) coefficient
functions in one-particle inclusive $e^+e^-$-annihilation. These results form
the basis for extracting fragmentation functions for light and heavy flavors
with NNLO accuracy that will be needed at the LHC and ILC.
The two-loop calculations have been performed in Mellin space based
on a new method, the main features of which we also describe
briefly.

\end{abstract}

\maketitle

\section{Introduction}
\label{sec:intro}
Observables with identified hadron(s) in the final state depend on
the fragmentation function(s) for these hadron(s). A
prominent example for the physics at hadron colliders is the
$p_T$-spectrum of $b$-flavored
mesons~\cite{Cacciari:2002pa,Cacciari:2003uh,Baines:2006uw},
\begin{equation}
\label{eq:bpt}
{d\sigma_B\over dp_T} \:\simeq\: \sum_{i,j,k} f_i\otimes
f_j\otimes {d\hat\sigma_{i,j\to k}\over dp_T}\otimes D_{k/B}
\, .
\end{equation}
Unlike the parton distributions (pdf's) $f_i$, the scale dependence
(evolution) of the fragmentation functions $D_{k/B}$ is in terms of
{\it time-like} splitting functions. As is well-known~\cite
{Curci:1980uw,Furmanski:1980cm,Floratos:1981hs},
starting from two loops (i.e., NLO) the
time-like and the space-like splitting functions of QCD differ.

To derive the time-like splitting functions to any order in
perturbation theory, one can exploit the universal factorization
property of the collinear singularities in some particular process,
for instance in $e^+e^-$-annihilation to hadrons,
\begin{equation}
\label{eq:e+e-}
e^+e^- \to f(p)+X \, ,
\end{equation}
where $f(p)$ can be any massless on-shell parton. The differential
cross-sections for the reaction~(\ref{eq:e+e-}) are known to
two loops~\cite{Rijken:1996vr,hep-ph/9609379,Rijken:1996ns,Mitov:2006wy},
permitting the extraction of the time-like NLO splitting functions and
NNLO coefficient functions.

The full three-loop calculation, needed for obtaining the NNLO
splitting functions is a problem of remarkable complexity due to,
among other reasons, the incomplete inclusiveness of the final state in
this process.
In contrast to the recent calculation of deep-inelastic scattering
(DIS) \mbox{\cite{Moch:2004pa,Vogt:2004mw,Moch:2004xu,Vermaseren:2005qc}},
one cannot apply the optical theorem here.
Progress in this direction is being made~\cite{Mitov:2005ps}
which will be discussed in Sec.~\ref{sec:epemfrag}.
Before we present in Sec.~\ref{sec:time-likeNNLO} an alternative
approach and first results for the time-like NNLO splitting functions.

\section{Time-like NNLO splitting functions}
\label{sec:time-likeNNLO}
The essence of our approach is to devise a relation
between the space- and time-like splitting functions in QCD capable
of predicting the quantities with time-like kinematics from the
known space-like ones~\cite{Moch:2004pa,Vogt:2004mw}.
This approach is a subject with interesting history starting with the
Gribov-Lipatov relation at leading order~\cite{Gribov:1972ri}. The NLO
calculations~\cite{Curci:1980uw,Furmanski:1980cm,Floratos:1981hs}
explicitly demonstrated the breaking of the Gribov-Lipatov relation
beyond the leading order, see also Ref.~\cite{Stratmann:1996hn}.
The interest in the subject has been re-emerging throughout the years
\cite{Broadhurst:1993ru,Brodsky:1995tb,Blumlein:2000wh},
and the space- or time-like two-loop results were independently
confirmed by several groups
\cite{Mitov:2006wy,Ellis:1996nn,Moch:1999eb,Kosower:2003np}.
New results beyond two loops, however, were obtained only very recently
\cite{Mitov:2006ic}.

Our starting point is the factorization of massless differential cross-sections.
Specifically, we consider in parallel the following two reactions:
\begin{eqnarray}
\label{eq:reactions}
{\rm DIS}\ :& f(p) +\gamma^* & \to \,\, X \, ,\\[1ex]
e^+e^-\ :& \hspace*{10mm} \gamma^* & \to \,\, f(p) + X \, ,
\nonumber
\end{eqnarray}
and the corresponding parton level observables
\begin{eqnarray}
\label{eq:observables}
{d\sigma\over dx}\;,& \displaystyle x ={-q^2\over 2 p \cdot q}\;,& q^2<0 \;, \\[1ex]
{d\sigma\over dz}\;,& \displaystyle z = {2 p \cdot q\over q^2}\;,& q^2>0 \;.
\nonumber
\end{eqnarray}

The above differential distributions contain singularities of
collinear origin. These, however, completely
factorize and can be absorbed into the definition of the
corresponding non-perturbative parton distribution or fragmentation
functions. Thus, they are completely specified in
terms of the kernels of the evolution equation.

The observables in Eq.~(\ref{eq:reactions}) (including $Z$-boson exchange)
are expressed in terms of three structure functions.
In DIS these are known as $F_1, F_2$ and $F_3$
while their counterparts in $e^+e^-$ are denoted as $F_T, F_L$ and
$F_A$. In the following we will consider only the relationship
between the functions $F_1$ and $F_T$, the others being similar.

We first take the bare (i.e. collinear singularities are not
factored out) structure function $F_1^{\,\rm b}$ expanded in
terms of the bare strong coupling $\als^{\,\rm b}$,
\begin{eqnarray}
\label{eq:F1b}
{\lefteqn{F_1^{\,\rm b} (\als^{\,\rm b}, Q^2) \; = \;}} \\
&&
  \delta(1-x) \:  + \: \sum_{l=1}^{\infty}
  \left( { \als^{\,\rm b\:\!} \over 4 \pi} \right)^l
  \left( Q^2 \over \mu^2 \right)^{-l\ep} F_{1,l}^{\,\rm b}
\:\: .
\nonumber
\end{eqnarray}
The terms in the above series are written as
\begin{eqnarray}
\label{eq:F1dec}
F_{1,1}^{\,\rm b}
     &\!=\!& 2 {\cal F}_1\,\delta(1-x) + {\cal R}_{\,1}  \\[1ex]
F_{1,2}^{\,\rm b}
     &\!=\!& 2 {\cal F}_2\, \delta(1-x)
           + \left({\cal F}_1\right)^2 \delta(1-x) \nn \\
          && + 2 {\cal F}_1 {\cal R}_{\,1} + {\cal R}_{\,2} \nn \\[1ex]
F_{1,3}^{\,\rm b}
     &\!=\!& 2 {\cal F}_3\, \delta(1-x)
           + 2 {\cal F}_1 {\cal F}_2\, \delta(1-x) \nn \\
          && + (\, 2 {\cal F}_2 + \left({\cal F}_1\right)^2 \,)\, {\cal R}_{\,1}
           + 2 {\cal F}_1 {\cal R}_{\,2} + {\cal R}_{\,3} \nn
\:\: .
\end{eqnarray}
This iterative decomposition into form-factor ($\cal F$) and real-emission
$(\cal R$) parts is done according to the $\ep$-dependence in the
soft limit, see Ref.~\cite{Moch:2005id}. It represent the closest
approach to a full decomposition into contributions from the separate
physical cuts possible on the basis of Refs.\
\mbox{\cite{Moch:2004pa,Vogt:2004mw,Moch:2004xu,Vermaseren:2005qc}}.

To predict the time-like quantities from their space-like
counterparts, we subject Eq.~(\ref{eq:F1dec}) to the following
analytical continuation: first, as can be easily seen from the
definitions~(\ref{eq:observables}) of the kinematical variables
in the two processes, one changes $x\to 1/x$ and $q^2\to - q^2$.
Accounting for the difference in the two phase-space measures, one
has to multiply the time-like expression by an overall factor of
$z^{1-2\epsilon}$ (see \cite{Mitov:2006wy} for more details). The
only subtle point in the analytic continuations is the treatment of
logarithmic singularities for $x \ra 1$,
cf. Ref.~\cite{Stratmann:1996hn}, starting with
\begin{equation}
\label{eq:l1xcnt}
  \ln (1-x) \;\ra\; \ln (1-z) - \ln z + {\rm i}\,\pi \:\: .
\end{equation}

After the continuation has been performed, we keep only the real
parts of the continued functions ${\cal R}_{\,l}$ and then
re-assemble the resulting expression analogously to
Eq.~(\ref{eq:F1dec}). This later result is identified with the
corresponding structure function $F_T^{\,\rm b}$ in $e^+e^-$ which,
according to the mass factorization, takes the form
\begin{eqnarray}
\label{eq:FTfact}
 F^{}_{T,1} &\!=\!& \!
   - \,\ep^{-1}\, P^{(0)} \: +\: c^{(1)}_{T} \: +\: \ep\, a^{(1)}_{T}
   \\
   &&\: +\: \ep^2\, b^{(1)}_{T} \: +\: \ep^3\, d^{(1)}_{T} \: +\: \ldots
   \, ,
\nn
\end{eqnarray}
and correspondingly for the coefficients $F^{}_{T,2}$ and $F^{}_{T,3}$ (see
Ref.~\cite{Mitov:2006ic} for their explicit expressions). The same
procedure applies to the other two structure functions $F_L^{\,\rm b}$
and $F_A^{\,\rm b}$.

The above described analytic continuation can in principle be
applied to any order in the strong coupling and to any order in the
expansion in~$\ep$. Indeed, by comparing to the known expressions at two
loops~\cite{Curci:1980uw,Furmanski:1980cm,Floratos:1981hs,%
Rijken:1996vr,hep-ph/9609379,Rijken:1996ns,Mitov:2006wy}
for the time-like splitting functions $P^{(1)}$
as well as the coefficient functions
$c^{(2)}$ and the new $a^{(2)}$ terms, we observe complete agreement
 -- which, beyond order $\ep^{-1}$, depends on the inclusion of the ${\rm i}\pi$
term in Eq.~(\ref{eq:l1xcnt}).
Moreover, all terms proportional to $\ep^{k},\ k\leq -2$
at three loops are also correctly predicted by the analytic
continuation carried out in this manner.

Despite the highly non-trivial predictions mentioned above, we
would like to stress the following point. Unlike the per-diagram
treatment of Refs.~\cite
{Curci:1980uw,Furmanski:1980cm,Floratos:1981hs,Stratmann:1996hn}, our
continuation relies on the space-like evaluation of Refs.\
\cite{Moch:2004pa,Vogt:2004mw,Moch:2004xu,Vermaseren:2005qc} based on
the optical theorem. Therefore, we have incomplete information on
the separate contributions from the various physical cuts, recall that
Eq.~(\ref{eq:F1dec}) does not represent a full decomposition
according to the number of emitted partons.
Thus one has to be prepared for some problem at the third order,
especially in the abelian ($C_F^{\,3}$) piece, related to $\pi^2$
contributions originating from phase space integrations over unresolved
regions.

There are two checks on the predicted three-loop time-like
splitting functions: their soft behavior $x\to 1$ and the vanishing
of a first moment $N=1$. The latter check shows a mismatch in
a term of the type $ \pi^2 C_F^3 (1+x^2)/(1-x) \ln^2 x$,
i.e., the analytic continuation returns an incorrect coefficient for
this term which has been restored by the $N=1$ constraint.

Clearly, one needs to devise a second and independent
confirmation of this corrected prediction. For this purpose we adopt
the approach of Dokshitzer, Marchesini and Salam~\cite{Dokshitzer:2005bf}.
There the evolution equations for either
parton distributions or fragmentation functions $f_{\sigma}^{\,\rm ns}$
are rewritten in a unified manner for
space-like ($\sigma=-1$) and
time-like ($\sigma=+1$) kinematics,
\begin{eqnarray}
  \label{eq:DMSevol}
  && {d \over d \ln Q^2} \: f_{\sigma}^{\,\rm ns} (x,Q^2) \; = \;  \\[1ex]
  && \int_x^1 {dz \over z} \: P^{\,\rm ns}_{\rm univ} \left( z,\als (Q^2) \right)
  \:  f_{\sigma}^{\,\rm ns} \Big( {x \over z},z^{\sigma}Q^2 \Big) \:\: ,
  \nn
\end{eqnarray}
with the universal splitting functions $P^{\,\rm ns}_{\!\rm univ}$
assumed to describe both the time-like and space-like cases.

One can easily work out the perturbative expansion of
Eq.~(\ref{eq:DMSevol}). The resulting expression for the difference
$P^{\,(1)\,\rm ns}_{\sigma=1}(x) - P^{\,(1)\,\rm ns}_{\sigma=-1} (x)$
at NLO agrees with Ref.~\cite{Curci:1980uw},
while the NNLO prediction coincides with the ($N=1$ corrected) result from our
analytical continuation.
It can be cast in the following very compact form
with $\otimes$ denoting the usual convolution,
\begin{eqnarray}
  \label{eq:dP2DMS}
  \delta\, P^{\,(2)\,\xi}(x) & = &
   2 \left\{ \Big[ \ln x \cdot \widetilde{P}^{\,(1)\,\xi} \Big]
              \otimes P^{\,(0)} \right.  \\
  &&  \left. + \Big[ \ln x \cdot P^{\,(0)} \Big]
              \otimes \widetilde{P}^{\,(1)\,\xi} \right\}
            \, ,
            \nn
\end{eqnarray}
for the non-singlet kernels ($\,\xi = +,\: -,\: \rm v$) with
\begin{equation}
\label{eq:Puni}
  2\,\widetilde{P}^{\,(n)\,\xi}(x) \; = \;
   P^{\,(n)\,\xi}_{\sigma=1}(x) + P^{\,(n)\,\xi}_{\sigma=-1}(x)
  \:\: .
\end{equation}

Obviously
Eq.~(\ref{eq:dP2DMS}) predicts the correct vanishing first moment for
$P^{\,(2)\,-}_{\sigma=1}(x)$ and, moreover, we can even make a prediction
for the fourth-order (N$^3$LO) difference $\delta P^{\,(3)\xi}$ of the (both
unknown) time-like and space-like non-singlet splitting functions
on this basis.
The difference can also be written in a form similar
to Eq.~(\ref{eq:dP2DMS}), see Ref.~\cite{Mitov:2006ic}
for explicit results.

\section{Hadron fragmentation in $e^+e^-$ at NNLO}
\label{sec:epemfrag}
Let us now turn to our method used for direct calculations
of higher-order QCD corrections in $e^+e^-$-annihilation~(\ref{eq:e+e-}).
In contrast to the preceding section where we have discussed
the analytic continuation in kinematic variables,
here we describe the derivation of the two-loop
corrections to the coefficient functions for light parton production.
They have been first calculated in
Ref.~\cite{Rijken:1996vr,hep-ph/9609379,Rijken:1996ns}, later confirmed
by the analytical continuation as outlined above and, finally,
re-derived through an independent direct two-loop calculation~\cite{Mitov:2006wy}.
The aim of the latter calculation was to check the earlier results of
Rijken and van Neerven and to derive the terms $a^{(2)}$
(defined as in Eq.~(\ref{eq:FTfact})).
The latter terms of ${\cal O}(\ep)$ were a useful check on predictions
of the analytical continuation at two loops, and will also be needed
in a future calculation of the three-loop corrections.

The calculation~\cite{Mitov:2006wy} was performed directly in Mellin space,
following the method of Ref.~\cite{Mitov:2005ps}, which consists of two
basic steps. First, one performs the Mellin transform before any
phase-space or virtual integration:
\begin{eqnarray}
\label{eq:sigma-N}
\sigma(N) &=& \int_0^1 dz z^{N-1}{d\sigma\over dz} \\[1ex]
&=& \int_0^1dz z^{n}\int d{\rm PS}^{(m)} \vert M\vert^2 \delta(z-f)\nonumber\\[1ex]
&=& \int d{\rm PS}^{(m)} \vert M\vert^2 f^{N-1} \, , \nonumber
\end{eqnarray}
where $f$ is some function of the external and/or integration momenta
that defines the variable $z$. If the measure $d{\rm PS}^{(m)}$
contains \mbox{$\delta$-functions}, they can be dealt with in the standard
way \cite{Anastasiou:2002yz}. From Eq.~(\ref{eq:sigma-N}) it is
clear that the effect of the Mellin transform is to introduce a new
propagator raised to an abstract power. The generalization to
multiple differential distributions involving a Mellin transform
in multiple variables is straightforward.

Second, one applies the reduction identities based on integration-by-parts (IBP)
to simplify the right-hand side of Eq.~(\ref{eq:sigma-N})
and to reduce it to master integrals.
The fact that one of the propagators is not a fixed integer but an abstract
parameter does not present any complications. This is because the
IBP identities can be cast in the usual form (where the powers of
all propagators are fixed integers) through a shift in the variable
corresponding to the power of the propagator $f$. The resulting
recurrence relations explicitly depend on $N$ and can be solved in
a standard way like, e.g., the algorithm of Laporta
\cite{Laporta:2001dd} as implemented in the {\sc Maple} package AIR of
Ref.~\cite{Anastasiou:2004vj}.

The master integrals resulting from the solution of the IBP
identities implicitly depend on $N$, too, and one can
derive for them difference equations in $N$.
To solve the latter we apply techniques used previously
in the context of DIS~\cite{Moch:2004pa,Moch:1999eb,Moch:2005id}
and perform the required expansions in $\ep$ with the packages
{\sc Summer}~\cite{Vermaseren:1998uu} and {\sc XSummer}~\cite{Moch:2005uc}.
By solving the difference equations one extracts the complete
dependence on the Mellin variable $N$ without the need for explicit
evaluations of any Feynman integral.
Finally, to fully specify the solution of a difference equation
of degree $\kappa$, one has to provide $\kappa$ initial conditions.
Here, the simplest option, which we have used in Ref.~\cite{Mitov:2006wy},
is the direct evaluation of the master integrals for $N=1,\dots ,\kappa$.

All but one of the master integrals in Ref.~\cite{Mitov:2006wy} satisfy
difference equations of first order, the other being of second order.
As pointed out in Ref.~\cite{Mitov:2005ps}, however,
not all of the integrals specifying the initial conditions
are independent, since the IBP reductions exhibit many additional
relations after fixing $N$, say to $N=1$.
Typically, for the initial conditions one has to explicitly evaluate
less objects than the number of master integrals.
In the case of Ref.~\cite{Mitov:2006wy}, we encountered six
real-real and respectively five real-virtual master integrals along with seven
independent initial conditions.

Finally, a particularly powerful feature of the method of
Ref.~\cite{Mitov:2005ps} is the fact that the master integrals depend
only on the process but are independent of the particular observable.
For that reason all integrals needed for the initial conditions
could be taken over from Ref.~\cite{DeRidder:2003bm}.
We confirmed those results and extended them to higher orders in $\ep$.

\section{Energy spectrum of $b$-quarks in $e^+e^-$}
\label{sec:b-quark}
The most important application of the time-like QCD splitting functions
is the description of hadron fragmentation in processes
sensitive to collinear radiation,
like in Eqs.~(\ref{eq:bpt}) or (\ref{eq:e+e-}).
For instance,
the process independent fragmentation functions $D_{k/B}$ of $b$-flavored
mesons are extracted from measurements in $e^+e^-$-annihilation.
Along with high-quality data, a consistent NNLO description of
light-flavor fragmentation requires the knowledge of the three-loop
time-like splitting functions and the two-loop coefficient
functions for the production of massless partons.
Additionally, within
the perturbative fragmentation function (PFF) formalism~\cite{Mele:1990cw}
for heavy-flavor fragmentation
one also has to supply initial conditions
now known to two loops~\cite{Melnikov:2004bm,Mitov:2004du}.
Flavor threshold crossing conditions may also have to be taken
into account~\cite{Cacciari:2005ry,Collins:1986mp}.

As an application of our results we can readily predict the
two-loop (NNLO) energy spectrum of massive $b$-quarks in $e^+e^-$-annihilation
based on the PFF formalism~\cite{Mele:1990cw}.
Now this formalism can be applied completely and consistently at two loops,
since all fermion-initiated components are presently
known~\cite{Melnikov:2004bm}.
The only two-loop component of the PFF not required in $e^+e^-$ at the
two-loop fixed-order level is the gluon-initiated
one~\cite{Mitov:2004du}.

The spectrum of a massive $b$-quark in $e^+e^-$-annihilation, up to power corrections
$\sim{\cal O}(m)$, reads:
\begin{eqnarray}
\label{eq:spectrumbmass}
{\lefteqn{ {d\sigma_b \over dz}(z,Q,m) =} } \\
&&
\sum_i{d\hat{\sigma}_i \over dx}(x,Q,\mu) \otimes {D}_{i/b}(x,m,\mu)
\, ,
\nonumber
\end{eqnarray}
where $\mu$ is the factorization scale and $Q$ represents the
characteristic hard scale of the reaction.

For the case of $e^+e^-$-collisions one usually takes $Q=\sqrt{s}$
and $m$ to be the pole mass of the heavy quark.
Inserting the explicit expressions for the coefficient functions and
the PFF's one can verify that indeed all dependence on the factorization scale $\mu$
completely cancels.

A particular feature of Eq.~(\ref{eq:spectrumbmass}) which we would like to point
out is the characteristic dependence $\sim 1/z$ at low $z$
that appears for a first time at two loops.
This is entirely due to the pure-singlet fermion emission
and similar to the well-known small-$x$ behavior of structure functions
in DIS~\cite{Moch:2004pa,Vogt:2004mw,Moch:2004xu,Vermaseren:2005qc}.

Detailed phenomenological studies of the energy spectrum
$d\sigma_b / dz$ of Eq.~(\ref{eq:spectrumbmass}) and related observables
will appear in a forthcoming publication \cite{MMb}.

\section{Conclusions}
\label{sec:conclusions}
In these proceedings we have reviewed recently derived results for
the non-singlet components of the three-loop (NNLO) time-like
splitting functions and the complete two-loop (NNLO) partonic
cross-sections for one-particle inclusive hadro-production in
$e^+e^-$-annihilation.

The former results~\cite{Mitov:2006ic} have been obtained by
utilizing an analytic continuation in the proper kinematic
variables in connection with the idea~\cite{Dokshitzer:2005bf} of a
universal splitting function governing the space- and time-like
parton evolution equations.

The latter results~\cite{Mitov:2006wy} provided a check on earlier
calculations by Rijken and van Neerven
\cite{Rijken:1996vr,hep-ph/9609379,Rijken:1996ns} and extended
their results to include higher order terms in $\ep$.
The derivation of these so far unknown terms of ${\cal O}(\ep)$
was essential in verifying predictions of the space-like to time-like
analytical continuation and will also be required in a future direct
three-loop evaluation of this process.

The derivation of the two-loop coefficient functions in $e^+e^-$
was a first application of a new method for evaluation of
differential distributions directly in Mellin
space~\cite{Mitov:2005ps}. This application demonstrated the
anticipated efficiency of the method both, in terms of solving the IBP
identities and, in the decrease of the number and the complexity of the
master integrals that require separate treatment.

The above results, together with the available two-loop
contributions~\cite{Melnikov:2004bm} to the PFF open the possibility
for studies of light and heavy hadro-production consistently at NNLO.
For example, the extraction of $b$-fragmentation functions at this
level of accuracy has important applications not only at a future ILC
but also at the LHC, for instance in measurements of the $p_T$-spectrum
of $B$-mesons or in precision determinations of the top-quark mass.\\

\end{document}